\documentclass[prb,aps,twocolumn,draft,showpacs]{revtex4}

\usepackage{graphicx}
\usepackage{graphpap}

\newcommand{\la}{\lambda}
\newcommand{\dq}{\delta q}
\newcommand{\dqE}{\delta q_\tE}
\newcommand{\dV}{\delta V}

\newcommand{\diag}{{\rm diag}}
\newcommand{\nond}{{\rm nond}}
\newcommand{\con}{{\rm con}}
\newcommand{\sing}{{\rm sing}}
\newcommand{\shell}{{\rm shell}}
\newcommand{\smooth}{{\rm smooth}}
\newcommand{\cost}{{\rm cost}}
\newcommand{\gain}{{\rm gain}}

\newcommand{\tF}{{\mbox{\tiny F}}}
\newcommand{\tE}{{\mbox{\tiny E}}}
\newcommand{\kFn}{k_{\tF\!,\,n}}
\newcommand{\kFnp}{k_{\tF\!,\,n'}}
\newcommand{\vFnu}{v_{\tF\!,\,\nu}}
\newcommand{\kFnu}{k_{\tF\!,\,\nu}}
\newcommand{\kEnu}{k_{\tE,\nu}}

\newcommand{\Jm}{ J_{m}}
\newcommand{\Km}{ K_{m}}

\begin{document}

\title{Scaling Theory of the Peierls-CDW in Metal Nanowires}

\author{D.~F.~Urban$^1$, C.~A.~Stafford$^2$, and Hermann~Grabert$^1$}

\affiliation{${}^1$Physikalisches Institut, Albert-Ludwigs-Universit\"at, D-79104 Freiburg, Germany \\
${}^2$Department of Physics, University of Arizona, Tucson, AZ
85721}

\date{\today}

\begin{abstract}

The Peierls instability in multi-channel metal nanowires is
investigated.  Hyperscaling relations are established for the
finite-size-, temperature-, and wavevector-scaling of the
electronic free energy. It is shown that the softening of surface
modes at wavevector $q=2\kFnu$ leads to critical fluctuations of
the wire's radius at zero temperature, where $\kFnu$ is the Fermi
wavevector of the highest occupied channel. This Peierls charge
density wave emerges as the system size becomes comparable to the
channel correlation length. Although the Peierls instability is
weak in metal nanowires, in the sense that the correlation length
is exponentially long, we predict that nanowires fabricated by
current techniques can be driven into the charge-density-wave
regime under strain.
\end{abstract}

\maketitle
\vskip2pc

\section{Introduction}

Already long ago, Fr\"ohlich \cite{Froehlich54} and Peierls
\cite{Peierls-book} pointed out that a one-dimensional metal
coupled to the underlying lattice is not stable at low
temperatures. Electron-phonon interactions lead to a novel type
of ground state with a charge-density wave (CDW) of wavevector
$2k_F$ (see Ref.\ \onlinecite{Gruener88} for a review). This
state is characterized by a gap in the single-particle excitation
spectrum, and by a collective mode with an associated charge
density \mbox{$\sim \rho_0+\rho_1\cos(2k_Fz)$}, where $\rho_0$ is
the unperturbed electron density of the metal.  Of particular
interest are {\em incommensurate} systems, where the period of
the CDW is not simply related to that of the unperturbed atomic
structure.  In that case, no long-range order is expected even at
zero temperature due to quantum fluctuations.

In contrast to the usual Peierls systems, metallic
nanowires\cite{Agrait03} are open systems with several
inequivalent channels, for which the theory has not yet been
developed. Interest in the Peierls transition in metal nanowires
has been stimulated by recent
experiments\cite{Yeom99,Ahn03,Ahn04,Ahn05,Snijders06} on nanowire
arrays on stepped surfaces. In these systems, interactions
between nanowires, mediated by the substrate, render
the system quasi-two-dimensional at low temperatures,
similar to the dimensional crossovers commonly observed in
highly-anisotropic organic conductors.\cite{Gruener88}  However,
individual free-standing metal nanowires\cite{Agrait03} represent
true (quasi)one-dimensional systems, in which the intrinsic
behavior of the Peierls-CDW can be studied.

Due to quantization perpendicular to the wire axis, electron
states in metal nanowires are divided into distinct channels,
which are only weakly coupled. Each channel has a quadratic
dispersion relation, and starts to contribute at a certain
threshold energy, i.e., the eigenenergy $E_n$ of the
corresponding transverse mode. This results in a sequence of
quasi-one-dimensional systems with different Fermi wavevectors
\begin{eqnarray}
    \kFn&=&\sqrt{\frac{2m_e}{\hbar^2}\left(E_F-E_n\right)}\,.
\end{eqnarray}
The channel Fermi wavevectors are generically {\em not
commensurate} with the underlying atomic structure in nanowires
with more than one open channel.

The Peierls instability is weak in metal nanowires,\cite{Urban03}
so that the system is close to the quantum critical point at
which the transition from Fermi liquid behavior to a CDW state
occurs. In the vicinity of the quantum critical point, the system
exhibits an additional length scale, the correlation length
\begin{eqnarray}
    \xi_{\nu}&=&\frac{\hbar\vFnu}{2\Delta_{\nu}}.
\end{eqnarray}
Here, the greek index $\nu$ labels the highest occupied channel,
in which an energy gap $2\Delta_{\nu}$ opens and
$\vFnu=({\hbar}/{m_e})\kFnu$ is the Fermi velocity of this
subband. Within a hyperscaling ansatz,\cite{Hertz76} the singular
part of the energy is expected to scale like
$E_{\sing}/L\sim\xi_{\nu}^{-1-z}$, where $L$ is the wire length,
and the dynamic critical exponent takes the value $z=1$.  Thus,
$E_{\sing}/L \sim \Delta_\nu^2/\hbar\vFnu$. Near the singular
point, we indeed find that the electronic energy is given by
\begin{eqnarray}
\label{eq:scalingrelation:Intro}
    \frac{E_{\sing}}{L}&\approx&\kappa_{\nu}\;
    \frac{\Delta_{\nu}^2}{\hbar\,\vFnu}
    \;\;\mbox{Y}\left(\xi_{\nu} \delta q,\frac{\xi_{\nu}}{L},
    \frac{\xi_{\nu}}{L_T}\right),
\end{eqnarray}
where $\mbox{Y}(x,y,z)\approx\ln(\mbox{max}\{x,y,z\})$ is a
universal and dimensionless scaling function, $\delta
q=(q-2\kFnu)$ is the detuning of the perturbation wavevector from
its critical value $2\kFnu$, $L_T=\hbar\,\vFnu/k_BT$ is the
thermal length at temperature $T$, and $\kappa_{\nu}=1\mbox{ or }
2$ is the degeneracy of the highest open channel $\nu$. This
universal scaling behavior is quite different from that of a
closed system with periodic boundary
conditions,\cite{Nathanson92,Montambaux98} where radically
different behavior was found for odd or even numbers of fermions.
Nonetheless, the correlation length $\xi$ was also found to
control the finite-size scaling of the Peierls transition in
mesoscopic rings.\cite{Montambaux98}

In this article, the quantum and thermal fluctuations of the
nanowire surface are calculated in a continuum
model,\cite{Zhang03} where the ionic background is treated as an
incompressible, irrotational fluid.  In contrast to the
semiclassical theory of Ref.\ \onlinecite{Zhang03}, which
exhibited critical surface fluctuations only at finite
temperature---due to the classical Rayleigh instability---the
present fully quantum-mechanical theory exhibits critical
zero-temperature surface fluctuations at wavevector $q=2\kFnu$.
In a finite system, these CDW correlations are found to grow in
amplitude as the wire length $L$ approaches a critical length
$L_c$ of order the correlation length $\xi_\nu$.  A similar
scaling is observed as the temperature is lowered, so that $L_T$
exceeds $\xi_\nu$.  Although $\xi_\nu$ is typically very large for
fully-equilibrated structures, consistent with the fact that
Peierls distortions have not yet been observed in multi-channel
nanowires, it is predicted that nanowires of dimensions currently
produced in the laboratory can be driven into the CDW regime by
applying strain.

This paper is organized as follows. Sec.\
\ref{sec:PeierlsInstability} summarizes the (standard) Peierls
theory for a one-dimensional metal with a half-filled band, and
extends it to multi-channel wires. A description of the
deformation of a nanowire through surface phonons is given in
Sec.\ \ref{sec:SurfacePhonons}. The correlation length $\xi$ is
introduced in Sec.\ \ref{sec:ScalingRelations}, and the scaling
relation (\ref{eq:scalingrelation:Intro}) is established. Sec.\
\ref{sec:Correlations} examines the critical surface
fluctuations, and consequences for different materials are
discussed. Finally, a summary and discussion are given in Sec.\
\ref{sec:Summary:Ch4}.

\section{The Peierls instability}
\label{sec:PeierlsInstability}

Consider the ground state of a one-dimensional linear chain of
atoms, with lattice constant $a$ and periodic boundary
conditions: In the presence of electron-phonon interactions, it is
energetically favorable to introduce a periodic lattice
distortion with period $\la=\pi/k_F$; this effect is known as the
Peierls instability.\cite{Peierls-book} The lattice distortion
opens up an energy gap $2\Delta$ in the electronic dispersion
relation at the Fermi level $E_F$, so that the total electronic
energy is lowered. If $\Delta\ll E_F$, then the gain in energy is
given by \cite{Peierls-book}
\begin{eqnarray}
\label{eq:Egain:gap:approx}
    \frac{E_{\gain}}{L}&\approx&
        \frac{1}{\pi}\,
        \frac{\Delta^2}{2E_F/k_F}\left[
        \log\!\left(\frac{\Delta}{4E_F}\right)-\frac{1}{2}
        \right]
        +{\mathcal{O}}\!
        \left(\!\frac{\Delta}{E_F}\!\right)^{\!4}\!\!\!.\quad
\end{eqnarray}
The size of the gap can be extracted from an energy balance: Let
the increase of the elastic energy due to the deformation be
given by $E_{\cost}/L=\alpha\,b^2$, where $b$ is the amplitude of
the distortion. On the other hand, $\Delta$ is defined through
the matrix element of the perturbation coupling
states with longitudinal wave vector $\pm k_F$;
it is linear in $b$ and we can set $\Delta=A\,b$. By finding the
minimum of $\delta E(b) = E_{\gain}+E_{\cost}$, we can derive the
optimal value of the distortion amplitude $b$ and from this
derive the size $\Delta$ of the gap,
\begin{eqnarray}
\label{eq:GapSize}
    \Delta&=&4E_F\;\exp\!\left(-2\pi\alpha E_F/A^2k_F\right).
\end{eqnarray}

The analysis of the Peierls instability in a one-dimensional
metal can be extended to the case of a multi-channel system.
While other instabilities of the Fermi liquid---induced by
electron-electron interactions---compete with the Peierls
instability in purely one-dimensional systems,\cite{Gruener88}
their importance decreases as the number of channels
increases,\cite{Matveev93} so that electron-electron interactions
can in a first approximation be neglected in multi-channel
nanowires. Moreover, electron-electron interactions are strongly
screened in s-orbital metal nanowires with three or more
conducting channels.\cite{Kassubek99,Zhang05} Hence, including
electron-electron interactions in the calculation will not lead
to any qualitative change (except in the limit of very few
conduction channels) whereas electron-phonon interactions
do.\cite{footnoteEE} We therefore consider only electron-phonon
coupling in this article.

For any given channel $n$, a perturbation of wavevector $q=2\kFn$
will open a gap $2\Delta_n$ at the Fermi surface in the energy
dispersion of this channel. The energy gain $E_{\gain,n}$ and gap
size $2\Delta_n$ are then given by Eqs.\
(\ref{eq:Egain:gap:approx}) and (\ref{eq:GapSize}), respectively,
with $k_F$ replaced by $\kFn$ and $E_F$ replaced by
$\hbar^2\kFn^2/2m_e$. The greatest effect will be seen close to
the opening of the highest occupied channel (i.e.\ $n=\nu$), where
the channel Fermi wavevector $\kFnu$ is small. Note that the same
perturbation (with $q=2\kFnu$) will also modify the dispersion
relations of lower-energy channels $n'$ with $E_{n'}<E_\nu$, but
due to the finite spacing of the threshold energies, this
modification will occur within the Fermi sea and there will be
little net effect.\cite{Urban03}


The standard Peierls theory uses periodic boundary conditions for
the wave functions. Its extension to the multi-channel case cannot
be directly applied to a metallic nanowire of finite length $L$.
The nanowire is part of a much larger system including the leads,
and the longitudinal wavevectors $k_n$ in the subbands are not
restricted to multiples of $\frac{2\pi}{L}$, as in the case of an
isolated system with periodic boundary conditions. Therefore a
perturbation of the nanowire with wavevector $q$ does not only
couple states $k_n$ and $k_n'$ that exactly obey $k_n=k_n'+q$.
Instead the state $k_n$ is coupled to a range of $k_n'$-states
proportional to $1/L$. The dispersion relation remains smooth
while, with increasing wire length, it develops a smeared-out
\emph{quasi-gap}, and only in the limit $L \rightarrow \infty$ do
we recover the Peierls result with a jump at $\kFn$.\cite{Urban03}

\section{Surface phonons}
\label{sec:SurfacePhonons}

We describe the wire in terms of the Nanoscale Free Electron Model
(NFEM),\cite{Stafford97a,Buerki05b} treating the electrons as a
Fermi gas confined within the wire by a hard-wall potential. The
ionic structure is replaced by a uniform (jellium) background of
positive charge, and we assume that this ionic medium is
irrotational and incompressible. The
approximations\cite{Kassubek99,Zhang03,Buerki05b} of the NFEM require strong
delocalization of the valence electrons, good charge screening,
and a spherical Fermi surface, conditions met in alkali metals
and, to a lesser extent, noble metals such as gold.

In this continuum model, the ionic degrees of freedom are
completely determined by the surface coordinates of the wire. Let
us consider an initially uniform wire of radius $R_0$ and length
$L$ which is axisymmetrically distorted. Its surface is given by
the radius function
\begin{eqnarray}
\label{eq:geometry}
    R(z,t)&=&R_0\biggl(1+\sum_q b_q(t) e^{iqz}\biggr),
\end{eqnarray}
where the time-dependent perturbation is written as a Fourier
series with coefficients $b_q(t)$.
Since $R(z,t)$ is real, we have $b_q=b_{-q}^*$ and we require
that the volume of the wire is unchanged by the deformation.
Other physically reasonable constraints are possible and will be
discussed in detail in Sec.\ \ref{sec:Correlations}.

The kinetic energy  of the ionic medium is given by\cite{Zhang03}
\begin{eqnarray}
\label{eq:Ekin:phonons}
    E_{kin}&=&
    L\sum_{q>0}m(q,{R_0})\left|{R_0}\frac{\partial b_q(t)}{\partial t}\right|^2.
\end{eqnarray}
Here, the \emph{mode inertia} $m$ is a function of wire radius and
phonon wavevector, and reads
\begin{eqnarray}
    m(q,R_0)&=&\rho_{ion}\cdot\frac{2\pi R_0 I_0(qR_0)}{q I_1(qR_0)}\;,
\label{eq:mode_inertia}
\end{eqnarray}
where $\rho_{ion}$ is the ionic mass density, and $I_0$ and $I_1$
are the modified Bessel functions of zeroth and first order,
respectively.\cite{footnote1} Considered as a function of $q$,
the mode inertia has a singularity $\sim 1/q^2$ at $q=0$, and is
monotonically decreasing, with  $m\sim 1/q$ for large $q$.

Within the Born-Oppenheimer approximation, the potential energy of
the ions is given by the grand canonical potential $\Omega$ of
the confined electron gas. A linear stability analysis of
cylindrical wires determining the leading-order change in
$\Omega$ due to a small perturbation was recently presented in
Ref.\ \onlinecite{Urban03}: $\delta\Omega$ is quadratic in the
Fourier coefficients $b_q$ of the deformation, and can be written
as
\begin{eqnarray}
\label{eq:Epot:phonons}
    \frac{\delta\Omega}{L}&=&\sum_{q\geq
    0}|b_q|^2\alpha(q,R_0,L,T)+{\cal
    O}\left(\frac{\lambda_F}{L}\right),
\end{eqnarray}
where the terms of order $\lambda_F/L$ include nondiagonal
contributions which can be neglected if the wire is long
enough.\cite{footnote4} The explicit analytical expression for
the \emph{mode stiffness} $\alpha$ is given in App.\
\ref{app:alpha}. Here, we are interested in the general behavior
of the mode stiffness as a function of the perturbation
wavevector $q$ for a given radius ${R_0}$, length $L$, and
temperature $T$: $\alpha(q)$ is a smoothly increasing function,
with a $L$- and $T$-dependent dip at $q=2\kFnu$, where $\nu$ is
the index of the highest open channel (see App.\
\ref{app:alpha}). Therefore, we formally split $\alpha$ into a
smoothly varying term and one containing the singular
contribution of channel $\nu$,
\begin{eqnarray}
\label{eq:modestiffness}
    \alpha=\alpha_\smooth + \alpha_\sing^{(\nu)}\;.
\end{eqnarray}
The smooth part can be thought of as the sum of a Weyl
contribution,\cite{Zhang03}
which describes the effects of surface tension $\sigma_s$
and curvature energy $\gamma_s$, plus an electron-shell correction\cite{Zhang03}
\begin{equation}
\alpha_\smooth = -2\pi R_0 \sigma_s + 2\pi R_0^2 (\sigma_s R_0 - \gamma_s)q^2 + \alpha_\shell,
\label{eq:alpha:smooth}
\end{equation}
whereas the singular part describes the onset of the Peierls
instability in channel $\nu$. At zero temperature, it is given
by\cite{Urban03}
\begin{eqnarray}
\label{eq:alpha:sing}
    \lefteqn{\alpha_\sing^{(\nu)}(q,{R_0},L)=
        -\frac{2m_e}{\hbar^2}\;\frac{4\kappa_\nu E_{\nu}^2}{\pi
        q}}\\
        &&
        \times\bigg[\ln\!\left|\frac{2\kFnu\!\!+\!q}{2\kFnu\!\!-\!q}\right|
        \!-\! \mbox{F}\Big(\!(2\kFnu\!\!+\!q)L\!\Big)
        \!+\!
        \mbox{F}\Big(\!|2\kFnu\!\!-\!q|L\!\Big)\bigg],\nonumber
\end{eqnarray}
where $\mbox{F}(x)=\mbox{Ci}(x)-\sin(x)/x$ and $\mbox{Ci}(x)$ is
the cosine integral function.\cite{footnote2} The finite
temperature mode stiffness is evaluated numerically by computing
the integral
\begin{eqnarray}
 \label{eq:DefAlphaT}
     \alpha(T)&=&\int dE
     \left(-\frac{\partial f}{\partial E}\right)\alpha(E)\;,
\end{eqnarray}
where $f(E)$ is the Fermi function and $\alpha(E)$ is obtained
from the zero temperature results by replacing $\kFnu$ by
$\kEnu\equiv[\frac{2m_e}{\hbar^2}(E-E_\nu)]^{-1/2}$.

\begin{figure}[bt]
    \begin{center}
    \setlength{\unitlength}{1mm}
    \begin{picture}(85,52)
       \put( 0,0){\includegraphics[width=5.1cm,draft=false]{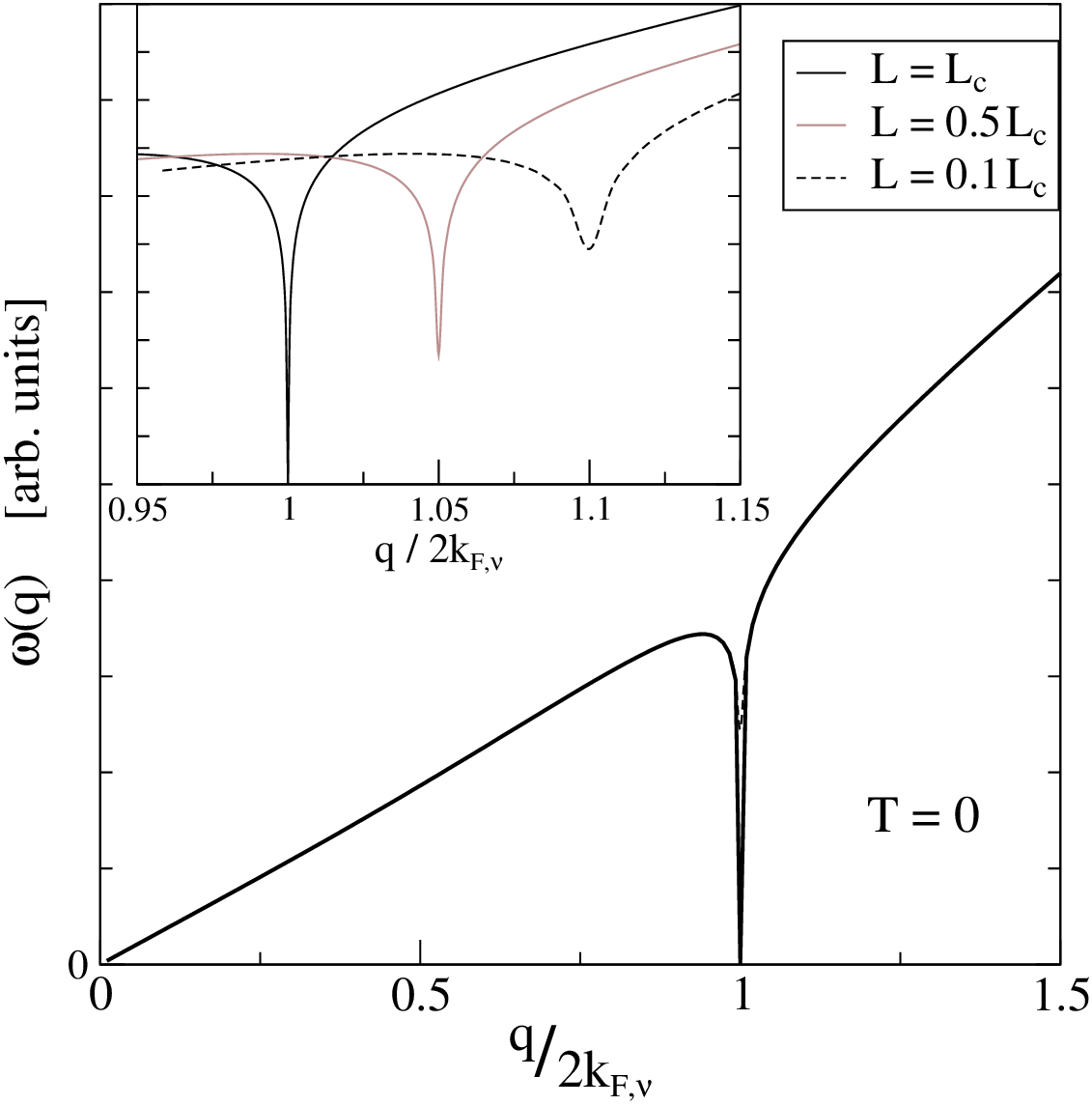}}
       \put(52,0){\includegraphics[width=3.2cm,draft=false]{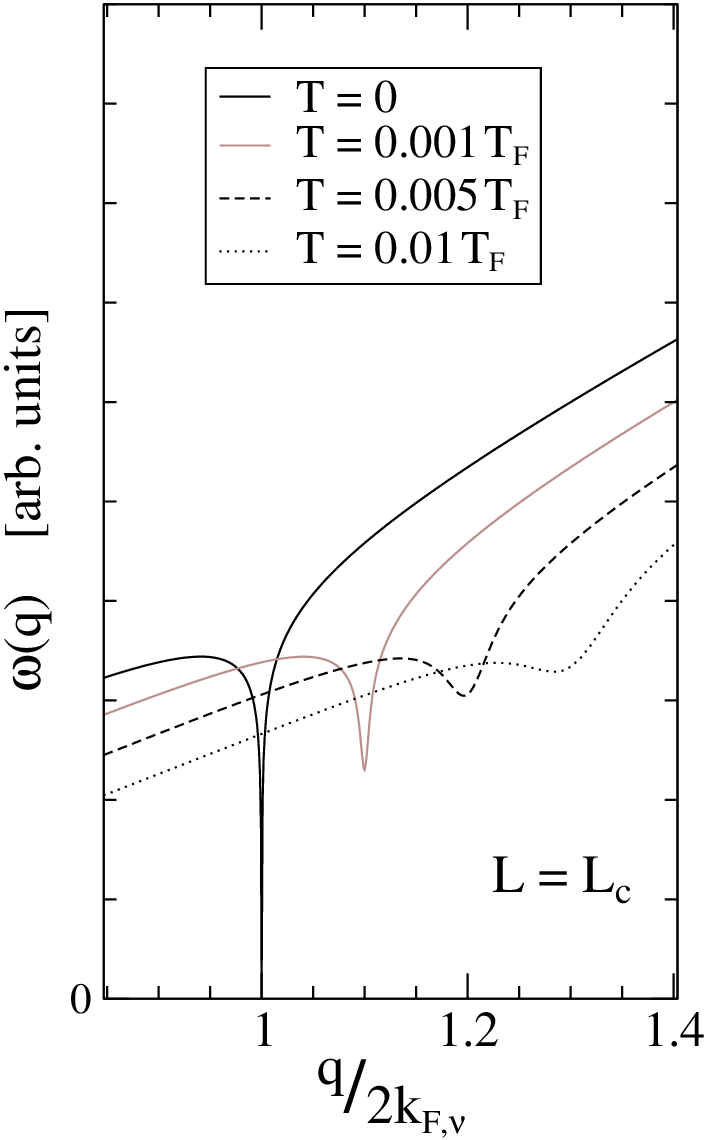}}
        \put(0,49){(a)}
        \put(51,49){(b)}
    \end{picture}
    \end{center}
    \vspace{-0.5cm}
    \caption[]{\label{Fig:omegaQ}(a) Surface phonon dispersion
    relation $\omega(q)$ for a wire of length $L=L_c$, where $L_c$ is
    the critical length (see text). Different
    wire lengths are compared in the inset showing the vicinity of
    $q=2\kFnu$. The curves are offset horizontally by $0.05$ units for
    clarity. (b) Temperature dependence of the minimum in
    $\omega(q)$ for a wire of length $L=L_c$ (curves are
    offset horizontally by $0.1$ units). The wire radius is ${R_0}=4.5k_F^{-1}$ for
    all curves, thus $L_c\sim6500k_F^{-1}$.
    }
\end{figure}

Combining the kinetic energy (\ref{eq:Ekin:phonons}) and potential
energy (\ref{eq:Epot:phonons}) yields a Hamiltonian  for surface
phonons, $H_{\mbox{\tiny
ph}}=\sum_q\hbar\omega_q\left(\hat{a}_q^{\dagger}\hat{a}_q+\frac{1}{2}\right)$,
with frequencies
\begin{eqnarray}
\label{eq:omegaPhonon}
    \omega(q,R_0,L,T)&=&\sqrt{\frac{\alpha(q,R_0,L,T)}{R_0^2\,m(q,R_0)}}.
\end{eqnarray}
These axisymmetric surface phonons correspond to the longitudinal acoustic mode of
the nanowire.
From Eqs.\ (\ref{eq:mode_inertia}), (\ref{eq:modestiffness}) and (\ref{eq:omegaPhonon}), we
infer that $\omega(q)$ is a smoothly increasing function of $q$, with a
dip at $q=2\kFnu$. As expected for acoustic phonons, it is linear
for small $q$. The $L$-dependent softening of the phonon modes
with wavevector $2\kFnu$ defines a \emph{critical length} $L_c$
for which $\omega(2\kFnu)=0$.

A plot of $\omega(q)$ for $L=L_c$ at zero temperature is shown in
Fig.\ \ref{Fig:omegaQ}\ (a). The inset shows a close-up of the
minimum and compares different wire lengths, where the curves are
horizontally offset for clarity. The temperature dependence of
$\omega$ is illustrated in Fig.\ \ref{Fig:omegaQ}(b),
concentrating on the vicinity of $q=2\kFnu$ (again, the curves
are horizontally offset). With increasing temperature, the dip in
$\omega$ disappears, and the curve becomes smoother.

\section{Scaling relations}
\label{sec:ScalingRelations}

The opening of the Peierls gap in subband $\nu$ introduces a new
energy scale, given by the gap $2\Delta_\nu$ for an infinitely
long wire. $\Delta_\nu$ is determined by the matrix element of the
perturbation coupling states with longitudinal wave vector $\pm
\kFnu$ and is linear in the distortion amplitude $b_q$. The
perturbation potential matrix element is calculated in App.\
\ref{app:matelement}, and we find that $\Delta_\nu=2E_\nu b_q$.

On the other hand, the energy cost for creating a surface
modulation with wavevector $q$ is determined by the smooth part
of the mode stiffness, so that $\delta
E_\cost/L=\alpha_\smooth(q)|b_q|^2$. Following the arguments of
Sec.\ \ref{sec:PeierlsInstability} we can now calculate the
length scale $\xi_\nu$, obtaining
\begin{eqnarray}
\label{eq:xi}
    \xi_{\nu}&=&\frac{\hbar\vFnu}{2\Delta_{\nu}}=\frac{1}{4\kFnu}
    \;\exp\!\left[\left(\!\frac{\hbar^2}{2m_e}\!\right)
    \frac{\pi\kFnu\alpha_{\smooth}}{2\kappa_{\nu}E_{\nu}^2}\right]\!,\quad
\end{eqnarray}
where we have used Eq.\ (\ref{eq:GapSize}). Introducing the
correlation length $\xi_\nu$ allows us to derive the finite-size-,
\mbox{temperature-,} and wavevector-scaling of the electronic free
energy near the critical point of the Peierls instability, given
by $q=2\kFnu$, $L\rightarrow\infty$, $T=0$, and weak electron--phonon coupling.

\emph{Finite-size scaling}---First we examine the mode stiffness
at zero temperature and for $q=2\kFnu$ as a function of wire length.
Starting from Eqs.\ (\ref{eq:modestiffness}) and
(\ref{eq:alpha:sing}), we take the limit $q\rightarrow 2\kFnu$ and
get
\begin{eqnarray}
    \alpha(L)\bigg|_{{q=2\kFnu}\atop{T=0\;\quad}}\!\!\!
    &=&\alpha_\smooth-\frac{4\kappa_\nu E_{\nu}^2}{\pi\hbar\vFnu}\,\left[
        \ln\!\left(4\kFnu L\right)-c_1\right],\qquad
\end{eqnarray}
where $c_1=1-\gamma_E+{\rm F}(4\kFnu L)\approx0.42$. Here
$\gamma_E\simeq0.577$ is the Euler-Mascheroni constant, and we
have used the fact that F$(x)\ll1$ for $x\gg1$. This expression
for $\alpha(L)$ can further be simplified by the use of
Eq.\ (\ref{eq:xi}), so that
\begin{eqnarray}
    \alpha(L)\bigg|_{{q=2\kFnu}\atop{T=0\quad\;}}
\label{eq:Lscaling:alpha}
    &=&\frac{4\kappa_\nu E_{\nu}^2}{\pi\hbar\vFnu}\,\left[
        \ln\!\left(\frac{\xi_\nu}{L}\right)+c_1\right].
\end{eqnarray}
This defines the critical length $L_c$ for which $\alpha$ takes
the value zero,
\begin{eqnarray}
\label{eq:Lcrit}
    L_c&\equiv&e^{c_1}\,\xi_\nu\;\approx\;1.52\,\xi_\nu\;.
\end{eqnarray}
Note that the critical length is of the same order of magnitude
as the correlation length.

\emph{Wavevector scaling}---Now let $\dq\equiv(q-2\kFnu)$ be the
detuning of the perturbation wavevector from its critical value
$2\kFnu$. At zero temperature and in the limit
$L\rightarrow\infty$, we expand the mode stiffness as a function
of $\delta q$ and obtain
\begin{eqnarray}
\label{eq:Qscaling:alpha}
    \alpha(\dq)\bigg|_{{L=L_c}\atop{T=0\;\,}}&\simeq&
        \frac{4\kappa_\nu E_{\nu}^2}{\pi\hbar\vFnu}\,
        \ln\!\left|\xi_\nu\dq\right|\,+\,{\cal{O}}(\dq)\;.
\end{eqnarray}
Again, the result was written in a compact form by the use of
Eq.\ (\ref{eq:xi}).

\emph{Temperature scaling}---Finally, we examine the effect of
finite temperature for a wire of infinite length and a
perturbation of wavevector $2\kFnu$. The main effect of finite
temperature is to smear out the Fermi surface, so that the
critical wavevector $q$ is detuned by $\dqE=2(\kFnu-\kEnu)$ with
$\kEnu\equiv[\frac{2m_e}{\hbar^2}\left(E-E_\nu\right)]^{-1/2}$.
Starting from the scaling behavior for finite $\dq$-detuning,
Eq.\ (\ref{eq:Qscaling:alpha}), we calculate $\alpha(T)$ for small
$T$ from Eq.\ (\ref{eq:DefAlphaT}) by linearizing
$E-E_F\approx\hbar \vFnu(\kEnu-\kFnu)$ and find
\begin{eqnarray}
\label{eq:Tscaling:alpha}
    \alpha(T)\bigg|_{{L=L_c\quad}\atop{q=2\kFnu}}&\simeq&
    \frac{4\kappa_\nu E_{\nu}^2}{\pi\hbar\vFnu}\;\left[
    \ln\!\left|\frac{\xi_\nu}{\hbar\vFnu\beta}\right|+c_2\right],
\end{eqnarray}
with a numerical constant $c_2=-\gamma_E+\ln\pi\approx0.57$.

Combining the three scaling relations for length, perturbation
wavevector, and temperature, Eqs.\ (\ref{eq:Lscaling:alpha}),
(\ref{eq:Qscaling:alpha}), and (\ref{eq:Tscaling:alpha}),
respectively, we prove Eq.\ (\ref{eq:scalingrelation:Intro}) for
the singular part of the electronic energy.\cite{footnote7}

\begin{figure}[bt]
    \begin{center}
        \includegraphics[height=5cm,draft=false]{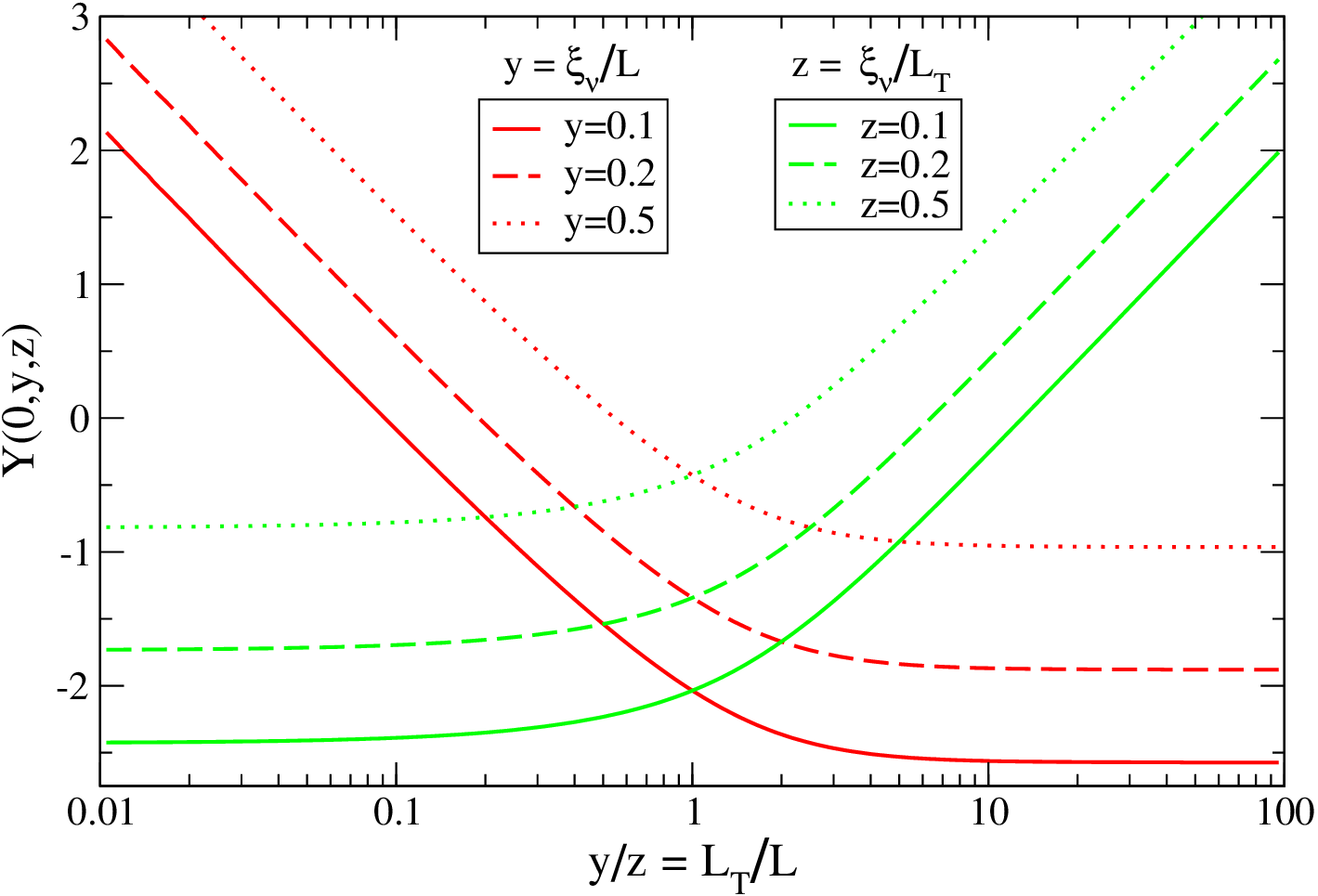}
    \end{center}
    \vspace*{-0.5cm}
    \caption[]{ \label{Fig:Yscaling:CrossOver} (color online) Cross-over from
    $T$-scaling to $L$-scaling: The scaling
    function $\mbox{Y}(x,y,z)$ is plotted as a function of the ratio $y/z=L_T/L$
    for fixed values of $y=\xi_\nu/L$ (red curves) and
    $z=\xi_\nu/L_T$ (green curves) at $x=\dq\xi_\nu=0$. For all curves $R_0=8k_F^{-1}$.
    }
\end{figure}

Finite temperature introduces a thermal length
$L_T=\beta\hbar\vFnu$. Changing $L_T$ has an equivalent influence
on the system properties as changing the length of the wire $L$.
Near the critical point, we observe a crossover between
finite-$T$ scaling and finite-$L$ scaling, depending on the ratio
$L_T/L$ (see Fig.\ \ref{Fig:Yscaling:CrossOver}). The smaller of
the two lengths dominates the behavior of the singular part of
the electronic energy.
As long as $L_T<L$, a change of $L$, even by orders of magnitude,
has only a small effect on Y, whereas the system is sensitive to
small changes in $L_T$ and shows $T$-scaling. The situation is reversed
for $L_T>L$: In this case, a change in temperature
results in only small changes of Y, whereas the singular part of
the energy depends strongly on $L$ and shows finite-length
scaling. These two different cases are illustrated in Fig.\
\ref{Fig:Yscaling:CrossOver} by the two sets of curves for
different fixed values of $L$ and $L_T$, respectively.

So far we have considered ideal nanowires without disorder.
Disordered structures exhibit an additional length scale, the electron
elastic mean free path $\ell$. The scaling theory we have derived
allows us to predict that the effect of disorder is to cut off
the logarithmic scaling of the Peierls-CDW instability, exactly
like the thermal length or wire length. We thus infer that the
length-dependent criterion for the emergence of the Peierls-CDW
in metal nanowires is given by
\begin{equation}
\label{eq:CompareLscales}
    L_c\lesssim L,\,L_T\,,\ell\;,
\end{equation}
where the critical length $L_c$ (Eq.\ \ref{eq:Lcrit}) is of order
the correlation length $\xi_\nu$. Increasing temperature leads to
a decreasing thermal length, and therefore destroys the
phenomenon at sufficiently high $T$. Increasing disorder has a
corresponding effect.

\begin{figure}[t]
    \begin{center}
            \includegraphics[height=5cm,draft=false]{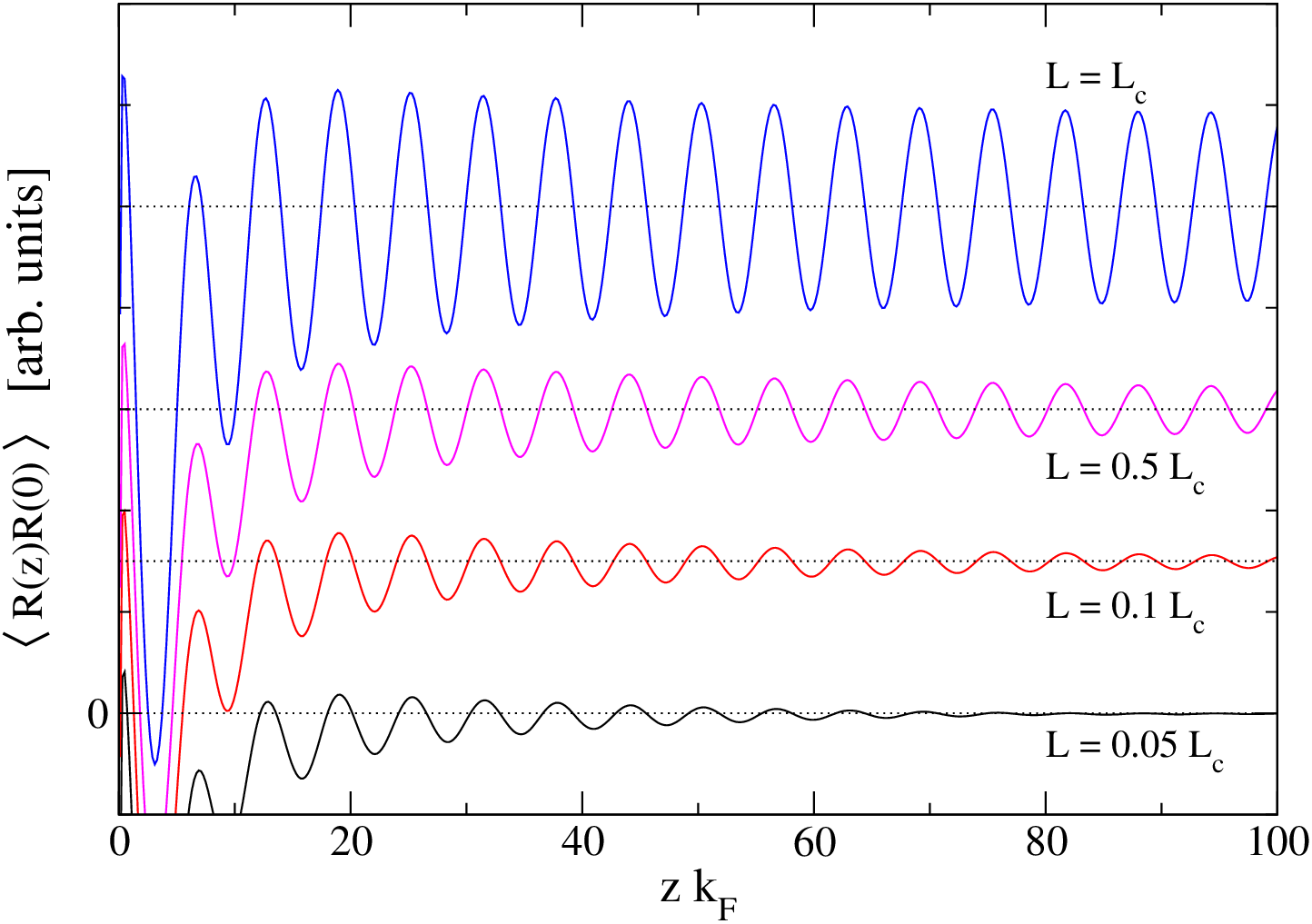}
    \end{center}
    \vspace*{-0.5cm} \caption[]{ \label{Fig:correlations}(color
    online) CDW correlations for various values of $L/L_c$ at $T=0$ for a wire
    with $R=4.42k_F^{-1}$. The critical length is $L_c\sim 1560k_F^{-1}$.
    Curves offset vertically for clarity.}
\end{figure}

\section{Surface fluctuations}
\label{sec:Correlations}

The softening of the surface phonon modes with wavevector
$q=2\kFnu$ leads to critical surface fluctuations. Given the mode
stiffness $\alpha$ and phonon frequency $\omega(q)$, the
fluctuations about the cylindrical shape are given
by\cite{footnote5}
\begin{equation}
\label{eq:correlation}
   \frac{\left\langle R(z)R(0)\right\rangle}{R_0^2}=\frac{1}{2\pi}\int_{-\infty}^{\infty}dq
   \frac{\hbar\omega(q)}{\alpha(q)}
   \left[\frac{1}{e^{\beta\hbar\omega(q)}-1}+\frac{1}{2}\right]e^{iqz}.
\end{equation}
Figure \ref{Fig:correlations} shows the correlations for
different wire lengths at zero temperature. Fluctuations with
$q=2\kFnu$ increase with increasing wire length. Note that these
CDW correlations may be pinned by disorder, or at the wire
ends.\cite{Gruener88} The correlations shown in Fig.\
\ref{Fig:correlations} are representative of regions far from an
impurity or wire end.

For large $z$, we can use a saddle point approximation to
estimate the integral in Eq.\ (\ref{eq:correlation}), and find
\begin{equation}
\label{eq:saddlepointapprox}
   \frac{\left\langle R(z)R(0)\right\rangle}{R_0^2} \propto
   \cos(2\kFnu z)\;K_0\!\left(\sqrt{12\log\!\left({L_c}/{L}\right)}\;\frac{z}{L}\right),
\end{equation}
where $K_0$ is the modified Bessel function of the second
kind\cite{footnote3} of order 0. Its asymptotic behavior is given
by\cite{Abramowitz-book}
\begin{eqnarray}
    K_0(x)&\sim&\left\{
    \begin{array}{cl}
    -\gamma_E-\log{x/2}  & \quad\mbox{for}\;x\ll1,\\
    \sqrt{\frac{\pi}{2x}}e^{-x} & \quad\mbox{for}\;x\gg1.
    \end{array}
    \right.
\end{eqnarray}
Depending on the ratio of the wire length $L$ to the critical
length $L_c$, we can distinguish two regimes: for $L\ll L_c$, the
prefactor of ${z}/{L}$ in the argument of $K_0$ in Eq.\
(\ref{eq:saddlepointapprox}) is large and the correlations decay
exponentially for sufficiently large $z$. On the other hand, if
$L\sim L_c$, our theory predicts a logarithmic decay of the
correlations. Note that the ratio $L/L_c$ determines the
crossover from a regime where the harmonic approximation about a
Fermi liquid is valid ($L<L_c$) to that of a fully-developed CDW ($L>L_c$). The
harmonic approximation, which we have used in our calculation,
breaks down at $L=L_c$, where the wire can no longer be treated
as a cylinder with small perturbations.

The correlation length $\xi$ is a material-specific quantity,
since it depends exponentially
on the smooth contribution (\ref{eq:alpha:smooth}) to the mode stiffness [cf.\ Eq.\
(\ref{eq:xi})].
As discussed in detail in Ref.\
\onlinecite{Urban06}, the material-specific surface tension and
curvature energy can be included in the NFEM through a
generalized constraint on the allowed deformations of the wire
\begin{equation}
\label{eq:constraint}
    {\cal N}=k_F^3{\cal V}-\eta_s\,k_F^2\,{\cal S}+\eta_c\,k_F{\cal C}\;=\;\mbox{const}.\;
\end{equation}
Here ${\cal V}$ is the volume
of the wire, ${\cal S}$ its surface area, and ${\cal C}$ its
integrated mean curvature.
The constraint $\partial{\cal N}=0$
restricts the number of independent Fourier coefficients in Eq.\
(\ref{eq:geometry}), and allows $b_0$ to be expressed in terms of
the other Fourier coefficients. This results in a modification of
the smooth part of the mode stiffness (see App.\ \ref{app:alpha}).
Through an appropriate choice of the dimensionless
parameters $\eta_s$ and $\eta_c$, the
surface tension and curvature energy can thus be set to the
appropriate values for any given material.\cite{Urban06}

\begin{figure*}[t]
    \begin{minipage}[c]{13.4cm}\hspace*{-6mm}
       \includegraphics[width=13cm,draft=false]{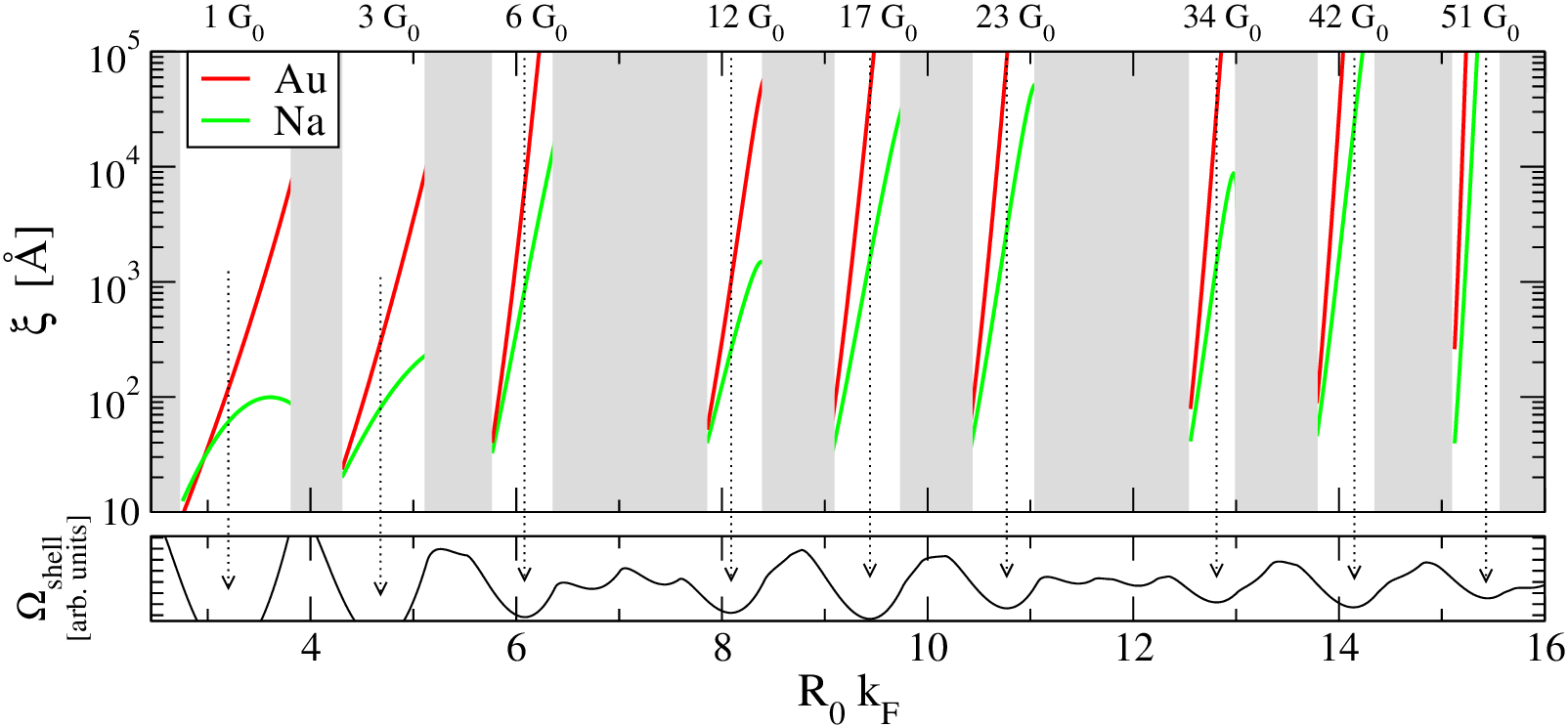}
    \end{minipage}\
    \begin{minipage}[c]{4.0cm}
       \caption[]{(color online) Upper panel: Correlation length $\xi_\nu$
       for Na and Au. For clarity, the plot is restricted to wires of
       so-called {\emph{magic radii}}, i.e.\ wires of conductance
       $G/G_0=1,3,6,12,...$, that were shown to be remarkably stable
       even when allowing symmetry breaking
       deformations.\cite{Urban03,Urban04} The lower panel shows the
       shell potential, and arrows mark the geometries with the longest
       estimated lifetime.\cite{Buerki05}
       }
       \label{Fig:xi}
    \end{minipage}
\end{figure*}

The upper panel of Fig.\ \ref{Fig:xi} shows the correlation length
calculated using the material parameters\cite{Tyson77} for Na and
Au. For clarity, the plot is restricted to the
cylindrical wires of so-called \emph{magic radii} that prove to
be remarkably stable even when allowing symmetry-breaking
deformations.\cite{Urban04,Urban06} These wires have conductance
values $G/G_0=1,3,6,12,17,23,34,42,51,...$, where $G_0=2e^2/h$ is
the conductance quantum.\cite{footnote6} Arrows point to the
positions of the most stable wires, defined as those with the
longest estimated lifetimes,\cite{Buerki05} which are near the
minima of the shell potential for straight cylindrical wires,
shown in the lower panel of Fig.\ \ref{Fig:xi}.

The maximum length of free-standing metallic nanowires observed
so far in experiments is that of gold wires produced by electron
beam irradiation of thin gold films in ultra high
vacuum,\cite{Kondo00} for which lengths of $L\sim3$--15\ nm are
reported. In those experiments, no sign of an onset of the
Peierls-CDW was seen, presumably indicating that $L<\xi$.
However, since the correlation length depends exponentially on
the wire radius, {\em we predict that nanowires of currently
available dimensions can be driven into the CDW regime by applying
strain}. A tensile force of order 1\ nN can change $\xi$ by
orders of magnitude, and thereby drive the system into the CDW
regime as soon as the condition (\ref{eq:CompareLscales}) is met.
Note that since $k_F$ is of order 1\ \AA$^{-1}$ for Na and Au,
the thermal length $L_T$ at room temperature is of order 100\
{\AA} for Na and twice as large for gold, comparable to the
lengths of the longest free-standing wires currently produced. By
contrast, the elastic mean-free path can be several times as
long\cite{Agrait03} and electron microscope images of gold
nanowires\cite{Kondo00} show perfectly regular and disorder-free
atomic arrangements. Surface roughness does not play a role in
the parameter regime we consider, where electron shell effects
dominate over ionic ordering. Thus CDW behavior should be
observable at room temperature in free-standing metal nanowires
under strain, in contrast to the behavior of
quasi-one-dimensional organic conductors,\cite{Gruener88} where
CDW behavior is observed only at cryogenic temperatures.

\section{Summary and Discussion}
\label{sec:Summary:Ch4}

In conclusion, we have presented a scaling theory of the
Peierls-CDW in multi-channel metal nanowires. Near the critical
point, scaling relations for the $L$-, $q$- and $T$-dependence of
the singular part of the free energy, which drives the Peierls
instability, were established. A hyperscaling ansatz was verified
and the universal scaling function was analyzed, which was found
to be logarithmic. The crossover from a regime where the harmonic
approximation about a Fermi liquid is valid ($L<L_c$) to that of
a fully-developed CDW ($L>L_c$) occurs at a critical length of
order the correlation length $\xi_\nu$, which is material dependent.
We predict that the Peierls-CDW should be observable at room
temperature in currently available metal nanowires under an
applied strain.

The critical length is shortest in materials whose surface
tension is small in natural units (i.e., in units of $E_F
k_F^2$).  Notable in this respect\cite{Tyson77} is Al with
$\sigma_s = 0.0018 E_F k_F^2$, some five times smaller than the
value for Au. Although Al is a multivalent metal, it has a very
free-electron-like band structure in an extended-zone scheme, and
thus may be treated within the NFEM, although the continuum
approximation is more severe. We thus predict that Al should be
an ideal candidate for the observation of the Peierls-CDW.

Our findings on the finite-size scaling of the Peierls-CDW in
metal nanowires are in contrast with previous theoretical
studies\cite{Nathanson92,Montambaux98} of mesoscopic rings: For
spinless fermions in a one-dimensional ring, the Peierls
transition was found to be suppressed for small systems when the
number of fermions is odd, but enhanced when the number is even.
Obviously, no such parity effect occurs in an open system, such
as a metal nanowire suspended between two metal electrodes; we
find that the CDW is always suppressed in nanowires with $L\ll
\xi_\nu$. Nonetheless, the correlation length $\xi$ was also
found to control the (very different) finite-size scaling in
mesoscopic rings.\cite{Montambaux98}

The finite-size scaling of the Peierls-CDW in metal nanowires is
similar to that of the metal-insulator transition in the
one-dimensional Hubbard model.\cite{Stafford93} In both cases,
there is no phase transition in an infinite system, because the
critical electron-phonon coupling and on-site electron-electron
repulsion is $0^+$ in each case. However, for parameters such
that the gap $2\Delta$ in an infinite system is sufficiently
small, the system is close to a quantum critical
point,\cite{Hertz76,Stafford93} and the crossover from
Fermi-liquid behavior to a Peierls-CDW or Mott insulator,
respectively, can be described within hyperscaling theory.

\section*{Acknowledgments}
We acknowledge the Aspen Center of Physics, where the final stage
of this project was carried out. This work was supported by the
DFG and the EU Training Network DIENOW (D.F.U. and H.G.) and by
NSF Grant No.\ 0312028 (C.A.S.).

\appendix

\section{Linear stability analysis}
\label{app:alpha}

This appendix gives details on the linear stability analysis
(cf.\ Ref.\ \onlinecite{Urban03}) which determines the
leading-order change in the grand canonical potential $\Omega$ of
the electron gas due to a small deformation of a cylindrical
nanowire. Since the nanowire is an open system connected to
macroscopic metallic electrodes at each end, it is naturally
described within a scattering matrix approach. The Schr\"odinger
equation can be expanded as a series in the perturbation, and we
solve for the energy-dependent scattering matrix $S(E)$ up to
second order. The electronic density of states can then be
calculated from
\begin{eqnarray}
    D(E) &=&\frac{1}{2\pi i}\;\mbox{Tr}\left\{
    S^{\dagger}(E)\frac{\partial S}{\partial E} -
        \frac{\partial S^{\dagger}}{\partial E}S(E)\right\},
\end{eqnarray}
where a factor of 2 for spin degeneracy has been included.
Finally, the grand canonical potential $\Omega$ is related to the
density of states $D(E)$ by
\begin{equation}
\label{eq:OmegaVonD}
    \Omega=-k_{B} T \int \!dE\; D(E) \;
    \ln\!\left[1+e^{-\frac{(E-\mu)}{k_{\mbox{\tiny B}}\,T}}
        \right],
\end{equation}
where $k_B$ is the Boltzmann constant, $T$ is the temperature, and
$\mu$ is the chemical potential specified by the macroscopic
electrodes. We find that the change $\delta\Omega$ due to the
deformation of an initially axisymmetric geometry in leading order
is quadratic in the Fourier coefficients $b_q$ of the deformation
(cf.\ Eq.\ \ref{eq:geometry}) and can be written as stated in
Eq.\ (\ref{eq:Epot:phonons}), defining the mode stiffness
$\alpha(q,R_0, L, T)$. It is convenient to decompose $\alpha$ into
three contributions,
$\alpha=\alpha_\diag+\alpha_\nond+\alpha_\con.$ The coupling
between channels mediated by the surface phonons determines
$\alpha_\diag$, coming from scattering into the same channel
(Eq.\ \ref{eq:alpha:diag}), and $\alpha_\nond$, coming from
scattering between different channels (Eq.\ \ref{eq:alpha:nond}):
\begin{widetext}
\begin{eqnarray}
\label{eq:alpha:diag} \alpha_\diag(q,R_0,L)&=&
        \frac{1}{\pi}\sum_{n}\Theta(E_F\!-\!E_{n})\left[
        12E_{n}\kFn-\frac{4E_{n}^2}{q}\left(
        \ln\left|\frac{2\kFn+q}{2\kFn-q}\right| - F((2\kFn+q)L) + F(|2\kFn-q|L)\right)
        \right]\qquad
\\
\label{eq:alpha:nond} \alpha_\nond(q,R_0,L)
    &=&-\frac{1}{\pi}\sum_{{n,n'}\atop{n\neq n'}}f_{n,n'}\Theta(E_F\!-\!E_{n})\;
    \left[16\kFn\frac{E_{n}E_{n'}}{E_{n}-E_{n'}}+\frac{4E_{n}E_{n'}}{q}
    \ln\!\left|\frac{q^2+E_{n'}-E_{n}+2q\kFn}{q^2+E_{n'}-E_{n}-2q\kFn}\right|\;+
\right.\nonumber\\
    &&\qquad\qquad\qquad
    +\Theta(E_F\!-\!E_{n'})\frac{4E_{n}E_{n'}}{q}\Big(
    F(|q-\kFn-\kFnp|L)-F((q+\kFn+\kFnp)L)\Big)
    \bigg]
\\
\label{eq:alpha:con}
\alpha_\con(q,R_0)&=&\frac{1}{\pi}\sum_{n}\Theta(E_F\!-\!E_{n})4E_{n}\kFn
    \frac{1+\left(\eta_c-\eta_s\,{R_0k_F}\right)
    \left({q}/{k_F}\right)^{\!2}}{1-{\eta_s}/({R_0k_F)}}.
\end{eqnarray}
\end{widetext}
Here $f_{n,n'}=1$ for two channels having the same azimuthal
symmetry and $f_{n,n'}=0$ otherwise. The function
$F(x)=\mbox{Ci}(x)-\sin(x)/x$ smoothens the logarithmic
divergences so that $\alpha_\diag$ and $\alpha_\nond$ are
continuous functions having minima of length-dependent depth at
$q=2\kFn$ and $q=\kFn+\kFnp$, respectively. The third
contribution (Eq.\ \ref{eq:alpha:con}) arises due to enforcing
the constraint (\ref{eq:constraint}) on allowed deformations.

\begin{figure}[t]
    \begin{center}
            \includegraphics[width=8.5cm,draft=false]{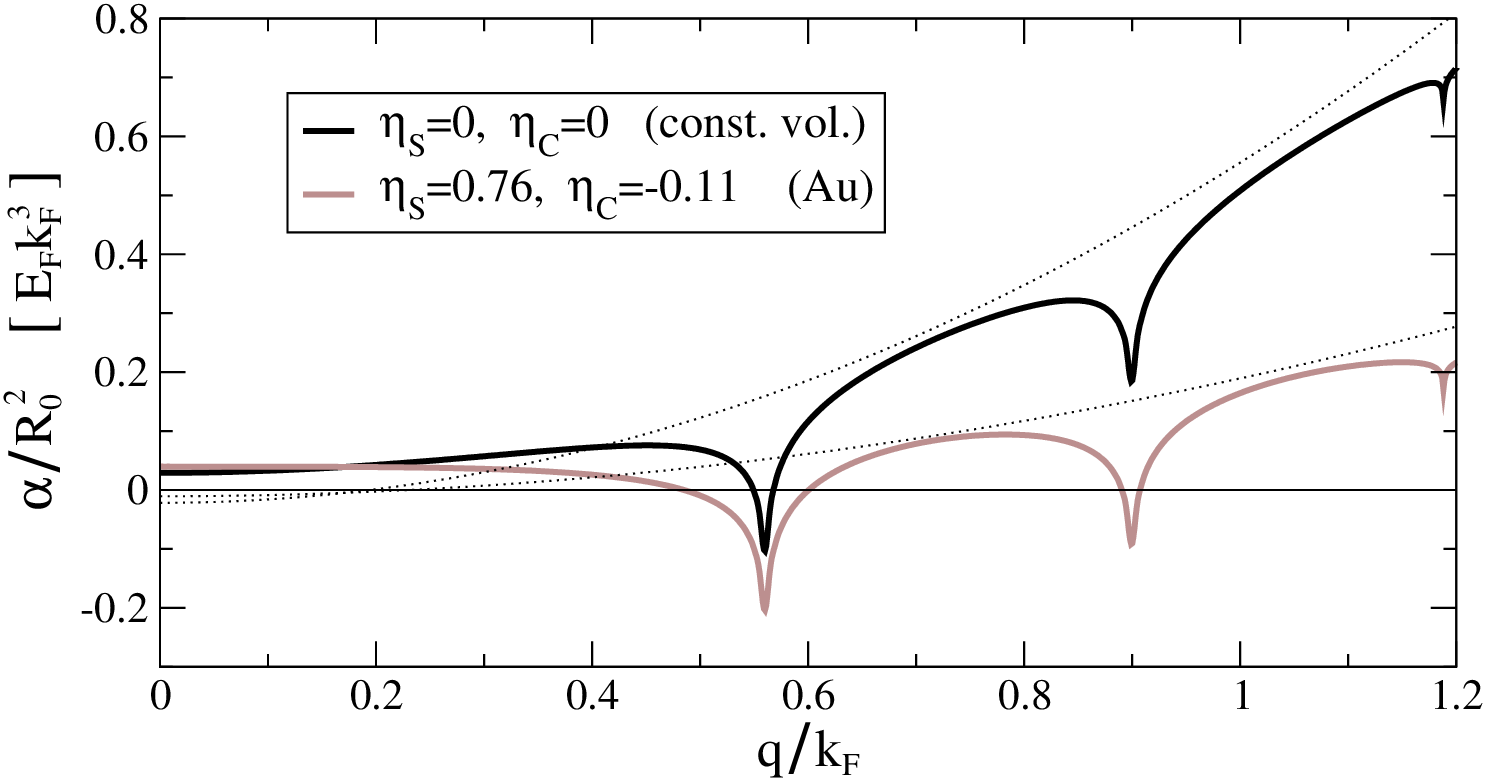}
    \end{center}
    \caption[]{(Color online) The mode stiffness $\alpha$
    computed from Eqs.\ (\ref{eq:alpha:diag})--(\ref{eq:alpha:con}) [solid curves].  The dashed curves
    show the Weyl approximation to $\alpha$ [first two terms on the r.h.s.\ of Eq.\ (\ref{eq:alpha:smooth})].
    The results for gold and for an idealized free-electron metal are shown; in each case, the
    radius is $R_0=5.75 k_F^{-1}$ and the length is $L=1000k_F^{-1}$.
    }
\label{fig:alpha_compare}
\end{figure}

The contribution to Eq.\ (\ref{eq:alpha:nond}) from evanescent
modes with $E_{n'}>E_F$ gives rise to the leading-order
$q$-dependence of $\alpha_\nond$, which is quadratic. This term
therefore essentially captures the change of surface and
curvature energy [cf.\ Eq.\ (\ref{eq:alpha:smooth})]. Figure
\ref{fig:alpha_compare} compares the mode stiffness $\alpha$,
computed from Eqs.\ (\ref{eq:alpha:diag})--(\ref{eq:alpha:con}),
to the Weyl approximation [first two terms on the r.h.s.\ of Eq.\
(\ref{eq:alpha:smooth})].  Here $R_0=5.75 k_F^{-1}$ and $L=1000
k_F^{-1}$.  Both the results for Au ($\eta_s=0.76$,
$\eta_c=-0.11$) and for a pure constant-volume constraint
($\eta_s=\eta_c=0$) are shown.  In both cases, the overall
$q^2$-dependence of $\alpha$ for large $q$ is evident, and
consequently, the minimum at $q=2\kFnu \approx 0.56 k_F$ is
deeper than the other minima.

When examining the softening of phonon modes, the overall
increase of $\alpha(q)$ with $q$ allows us to concentrate on the
$q_0\equiv 2\kFnu$ mode, where $\nu$ labels the highest open channel.
This mode will always be the dominant one. In general, the mode
stiffness also shows less-pronounced minima at other values $q>q_0$, due
to the lower-lying channels, but these contributions are less singular
than the $q_0$ mode.
It is therefore possible to approximate $\alpha$ by a smooth
contribution
$\propto q^2$ plus the explicit
terms of Eq.\ (\ref{eq:alpha:diag}) for the highest open channel.

\section{Perturbation potential matrix elements}
\label{app:matelement} Consider free electrons confined within a
cylindrical nanowire of radius $R_0$ and length $L$ by a step
potential $V(r)=V_0\theta(r-R_0)$, where $\theta$ is the step
function. The transverse eigenfunctions in polar coordinates are
given by
$\psi^\perp_{mn}(r,\varphi)=(2\pi)^{-\frac{1}{2}}e^{im\varphi}\chi_{mn}(r)$,
where the radial function $\chi_{mn}(r)$ reads
\begin{equation}
    \chi_{mn}
        =N_{mn}\!\left\{\!
        \begin{array}{ll}
        \Jm\!\big(\sqrt{E_{mn}}\,r\big), &r<R_0,\\
        \frac{\Jm\!\big(\!\sqrt{\!E_{mn}}\,R_0\big)\Km\!\big(\sqrt{V_0\!-\!E_{mn}}\,r\big)}{\Km\!\big(\!\sqrt{V_0-E_{mn}}\,R_0\big)}\,
        , &r>R_0.
        \end{array}\right.
\end{equation}
Here $N_{mn}$ is a normalization factor, $\Jm$ is the Bessel
function of order $m$, $\Km$ is the modified Bessel function of
the second kind of order $m$, and for simplicity of notation, we
use the convention ${\hbar^2}/{2m_e}=1$. The transverse
eigenenergies $E_{mn}$ are determined by the continuity of
$\partial_r\chi_{mn}/\chi_{mn}$ at $r=R_0$. An axisymmetric
perturbation of the wire changes the confinement potential by
\begin{equation}
    \dV(r,z)=V_0\big[\theta(r-R_0-\delta
    R(z))-\theta(r-R_0)\big],
\end{equation}
where the variation in radius is given by $\delta
R(z)/R_0=\sum_qb_qe^{iqz}$ and the perturbation wavevectors $q$
are restricted to integer multiples of $2\pi/L$. We expand the
matrix elements of $\delta V$ with respect to the unperturbed
eigenfunctions $\Psi_{mn k}(r\!,\varphi,z)=
L^{-1/2}e^{ikL}\psi^\perp_{mn}(r,\varphi)$ to first order in
$b_q$ and get
\begin{eqnarray}
    \langle mn
    k\vert\,\dV\vert\,\bar{m}\bar{n}\bar{k}\rangle&\simeq&
    -\delta_{m\bar{m}}V_0R_0\,{\chi}_{mn}(R_0){\chi}_{\bar{m}\bar{n}}(R_0)\nonumber\\
    &&\times\sum_qb_q\frac{1-e^{i(\bar{k}-k+q)L}}{i(\bar{k}-k+q)L}\,,
\end{eqnarray}
Taking the limit $V_0\rightarrow\infty$ we get
\begin{eqnarray}
    \lim_{V_0\rightarrow\infty}\langle mn
    k\vert\,\dV\vert\,m\bar{n}\bar{k}\rangle&\simeq&
    2\sqrt{E_{mn}E_{m\bar{n}}}\nonumber\\
    &&\times\sum_qb_q\frac{1-e^{i(\bar{k}-k+q)L}}{i(\bar{k}\!-\!k\!+\!q)L}.\qquad
\end{eqnarray}
In the limit of $L\rightarrow\infty$ we have
\begin{equation}
    \lim_{{V_0\rightarrow\infty}\atop{L\rightarrow\infty}}\langle mn
    k\vert\,\dV\vert\,m\bar{n}\bar{k}\rangle\simeq-
    2\sqrt{E_{mn}E_{m\bar{n}}}\;b_q\,\delta_{\bar{k},k-q}
\end{equation}
and recover a coupling of states with $\bar{k}=k+q$, only.

\vspace*{-2mm}

\bibliography{Urban06b}

\begin{thebibliography}{34}
\bibitem[{foo({\natexlab{a}})}]{footnoteEE}
\bibinfo{howpublished}{In that respect the situation is considerably different
  from, e.g., quasi-one-dimensional organic conductors where interactions lead
  to important qualitative effects.}
\bibitem[{foo({\natexlab{b}})}]{footnote1}
\bibinfo{howpublished}{The modified Bessel functions (of the first kind)
  $I_n(x)$ are solutions of the differential equation $x^2y''+xy'-(x^2+n^2)y=0$
  that are regular at $x=0$.}
\bibitem[{foo({\natexlab{c}})}]{footnote4}
\bibinfo{howpublished}{Note that the mode stiffness $\alpha(q,R_0, L, T)$
  appearing in Eq.\ (\ref{eq:Epot:phonons}) essentially determines the
  stability of the nanowire: if $\alpha$ is positive for all values of the
  perturbation wavevector $q$, all small deformations of the cylindrical
  geometry lead to an increase of $\Omega$ and are therefore suppressed.}
\bibitem[{foo({\natexlab{d}})}]{footnote2}
\bibinfo{howpublished}{The cosine integral function is defined as
  $\mbox{Ci}(x)=-\int_x^{\infty}dt\;\cos(t)/t$.}
\bibitem[{foo({\natexlab{e}})}]{footnote7}
\bibinfo{howpublished}{Hereby we need to remember that
  $E_\sing/L=\alpha\cdot|b_{2\kFnu}|^2$ and
  $\Delta_\nu^2=4E_\nu^2|b_{2\kFnu}|^2$.}
\bibitem[{foo({\natexlab{f}})}]{footnote5}
\bibinfo{howpublished}{This expression can be derived by using ${R_0}
  b_q(t)=\sqrt{\hbar/2m\omega
  L}\big(\hat{a}_qe^{-iwt}+\hat{a}_{-q}^{\dagger}e^{iwt}\big)$, which relates
  the deformation amplitude to the phonon operators, and by assuming thermal
  occupation numbers. Furthermore, we have approximated the sum over $q$ by an
  integral.}
\bibitem[{foo({\natexlab{h}})}]{footnote6}
\bibinfo{howpublished}{For completeness, the whole sequence of magic radii is
  included in Fig.\ \ref{Fig:xi} although it is clear that the continuum model
  used for our calculations is at its limits of validity for a wire of
  conductance $1G_0$.}
\end{thebibliography}

\end{document}